\documentstyle[prb,aps,epsfig,graphicx,tabularx, multicol]{revtex}
 
\title{Charge fluctuation induced dephasing in a gated mesoscopic interferometer}
 
\author{
Georg Seelig$^{*}$
and Markus B\"uttiker$^{\dagger}$
}
 
\address{ D\'epartement de Physique Th\'eorique,
 Universit\'e de Gen\`eve,
CH-1211 Gen\`eve 4, Switzerland}
 
\date{\today}

\newcommand{\ovl}{\overline}
\newcommand{\hw}{\hbar\omega}
\newcommand{\mybeginwide}{
    \end{multicols}\widetext
    \vspace*{-0.2truein}\noindent
    \hrulefill\hspace*{3.6truein}
}
\newcommand{\myendwide}{
    \hspace*{3.6truein}\noindent\hrulefill
    \begin{multicols}{2}\narrowtext\noindent
}
 
\begin{document}
\maketitle
\bigskip
 
\begin{abstract}
The reduction of the amplitude of Aharonov-Bohm oscillations in a 
ballistic one-channel mesoscopic interferometer due to charge fluctuations 
is investigated. 
In the arrangement considered the interferometer has four
terminals and is coupled to macroscopic metallic side-gates. 
The Aharonov-Bohm oscillation amplitude is calculated as a function 
of temperature and the strength of coupling between the ring and 
the side-gates. The resulting dephasing rate is linear in temperature
in agreement with recent experiments. Our derivation emphasizes 
the relationship between dephasing, ac-transport and charge fluctuations.

\end{abstract}
 
%\pacs{PACS numbers: 73.20.Dx Electron states in low-dimensional
%structures (superlattices, quantum well structures and multilayers)
%85.30.Vw Low-dimensional quantum devices
%(quantum dots, quantum wires etc.)
%03.67.Lx Quantum computation}
 
\begin{multicols}{2}                            
\narrowtext
 
\section{Introduction}
 
\begin{figure}
\includegraphics[scale=0.4]{{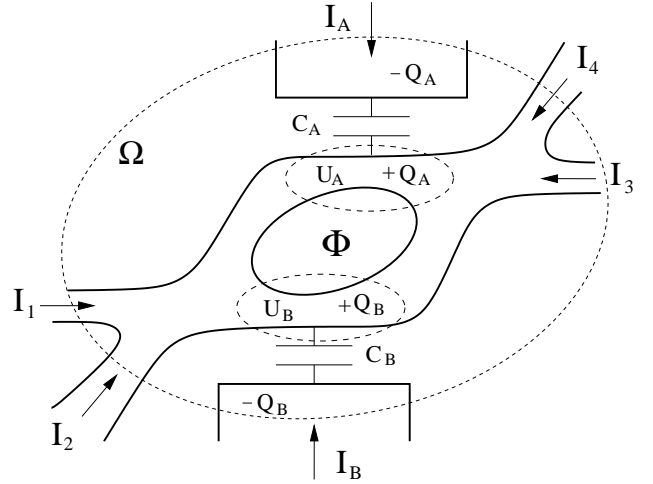}}
%\centerline{
%\epsfxsize8.5cm
%\epsffile{mzi_fig3.eps}}
\vskip 0.3cm
\caption{\label{mzi_figure}
 The figure shows the four-terminal AB-ring threaded by a magnetic flux. 
The two arms of the ring are each coupled to a side-gate via a 
capacitance $C_G$  ($G=A,B$). We consider junctions which are 
perfectly transmitting 
and divide the incoming current into the upper and lower 
branches 
of the ring. The system then is the electronic equivalent 
of an optical Mach-Zehnder interferometer (MZI). The total 
charge in a Gauss  sphere $\Omega$ drawn around the system 
of gates and ring is assumed to be zero, implying that current 
in the system is conserved. 
It is assumed that each arm is 
characterized by a single potential $U_A$ (or  $U_B$)
and the charge  $+Q_A$ (or  $+Q_B$).} 
\end{figure}

Dephasing processes suppress quantum mechanical
interference effects and generate the transition
from a microscopic quantum coherent world in which interference is
crucial to a macroscopic world characterized by the absence of (quantum) 
interference effects. Mesoscopic systems are neither entirely 
microscopic nor macroscopic
but at the borderline between the two. Clearly, therefore,
dephasing processes play a central role in
mesoscopic physics \cite{aak,chsm,sai,loss,blant}.
At low temperatures, it is thought that the predominant
process which generates dephasing, are
electron-electron interactions \cite{aak,chsm,blant}.
In this work, we investigate a 
ballistic Aharonov-Bohm (AB) interferometer, in which electrons
(in the absence of interactions)
are subject only to forward scattering processes (see Fig.~(\ref{mzi_figure})).

Our work is motivated by the following questions: 
A mesoscopic conductor connects two or more electron reservoirs:
inside a reservoir screening is effective and electron interactions are
of little importance. In contrast, inside the mesoscopic structure
screening is poor and interactions are important. Thus
the process of a carrier entering or leaving the mesoscopic
structure is essential. We ask how this process affects
the dephasing. In standard treatments of dephasing the conductor
is considered to be charge neutral and the elementary excitations
are electron-hole scattering processes. In a finite size mesoscopic
conductor, we can, however, have a hole in a reservoir and only an
additional electron in the conductor, or we can have an electron inside
the conductor and a hole on a nearby capacitor (see Fig. \ref{mzi_figure}).
As a consequence the conductor is charge neutral only when its surroundings
are taken into account (reservoir banks, nearby capacitors).  
    
A second question we seek to answer is the following: instead of
calculating a dephasing rate it is desirable to find a way
to directly evaluate the quantity of interest (here the conductance).
In small mesoscopic systems the dephasing rate might be a sample
specific quantity \cite{anka} and there would be little justification
in using an ensemble averaged dephasing rate even if we are
interested only in the ensemble averaged conductance. 
Clearly to answer such conceptual questions it is useful to have 
a model which is as simple as possible. 
In this paper we theoretically investigate dephasing of 
AB-oscillations in a ballistic ring with a single transport channel. 
The one-channel limit is of actual experimental interest 
(see e.g.  Refs.~\onlinecite{casse,hansen,hansen2}). Our idealized 
setup consists of an AB-ring with four terminals, the arms 
of the ring being capacitively coupled to lateral gates 
(see Fig.~(\ref{mzi_figure})). For a recent experiment 
on a ballistic (two-terminal) AB-ring with lateral gates 
coupled to both arms see Ref.\onlinecite{krafft}.
The structure we examine has no closed orbits: 
As a consequence in its equilibrium state it exhibits 
no persistent current and in the transport state 
there is no weak localization correction to the conductance.  
It exhibits, however, an Aharonov-Bohm effect \cite{ab} 
due to superposition of partial waves in the out-going final 
quantum channel. In fact this is the 
situation discussed in the original work of 
Aharonov and Bohm \cite{ab}. It is also sometimes 
assumed in mesoscopic physics without a detailed specification 
of the conditions (multi-terminal geometry, absence of 
backscattering)
which are necessary for interference 
to appear only in the final outgoing channel. 
The system investigated here is the electric analog of an 
optical interferometer in which the path is divided 
by forward scattering only. 
An example of such an arrangement is the Mach-Zehnder interferometer (MZI). 
In the MZI the sample specific AB oscillations are a consequence of 
superposition in the outgoing scattering channel only. 
We calculate the effect of internal potential fluctuations on the 
linear response dc-conductance in the ring as a function 
of the strength of the  coupling between ring and gates 
and of temperature. 
In this approach the conductance and the dephasing 
time are not calculated 
separately but 
a dephasing time will appear in the expression for 
the conductance in a natural way. It quantifies the 
degree of the attenuation of AB-oscillations due to 
randomization of the phases of the electrons going 
through the ring (as opposed to attenuation due to 
thermal averaging). From our calculations we find 
that the coherent part of the conductance is diminished 
by a factor $1-\tau/\tau_\phi \approx {\rm exp}(-\tau/\tau_\phi)$ 
relative to the ideal case due to  temperature and coupling 
to the gates. Here $\tau$ is the traversal time for going 
through one arm of the ring. In the temperature regime 
$0\ll kT\ll \Delta E$ ($\Delta E$ is the sub-band threshold) 
addressed in our calculations we find a dephasing rate $\tau^{-1}_\phi$
{\it linear} in temperature. In an experiment on a two-terminal 
AB-ring, a dephasing rate linear in temperature was 
recently measured by Hansen et al. ~\cite{hansen}. 

In our model the dephasing is due to inelastic 
scattering of electrons from charge fluctuations in the arms of the 
ring. We treat the gates as macroscopic entities with perfect screening.
The carrier dynamics in the gates is irrelevant for the discussion presented
here. They do not represent an external bath or dephasing agent. 
The irreversible source necessary for dephasing is 
given by the electron dynamics of the ring itself: the phase and energy of a 
carrier exiting into a contact are unrelated to the phase and energy of a
carrier entering the sample.

Our approach is similar in spirit to the one used in 
Ref.~\onlinecite{am} where dephasing due to charge fluctuations was 
discussed for two coupled mesoscopic structures.
To simplify the discussion we assume 
that it is possible to draw a Gauss sphere around the system 
of gates and ring such that all electric field lines 
emanating within the sphere also end in it 
(see Fig.~(\ref{mzi_figure})). This implies that the total 
charge in the sphere is zero at any time. However, it is 
possible to charge up one part of the system (an arm of the ring) 
relative to another part (the nearby gate) creating {\it charge dipoles}. 
Charge fluctuations 
in the arms of the ring lead to fluctuations of the 
effective internal potentials. Electrons going 
through the ring are exposed to these potential 
fluctuations and scatter inelastically. The strength 
of the coupling between the gates and the arms determines 
the amount of screening and thus the strength of the 
effective electron-electron interaction. When the 
capacitance $C$ between arm and gate becomes very large, 
the Coulomb energy $e^2/C$ of the system goes to zero and 
gate and arm are said to be decoupled.
  
The main goal of this work is the calculation of 
the dc-conductance of the ring taking into account 
the effects of the gate-mediated interactions. 
Applying the ac-scattering approach of Ref.\onlinecite{ap1} we start 
by calculating the dynamic conductance matrix of our system. 
From the real (dissipative) part of the dynamic conductance 
matrix element relating the current in one of the gates to 
the voltage applied to the same gate we can find the spectrum 
of the equilibrium potential fluctuations in the nearby arm of 
the ring via the fluctuation dissipation theorem. When calculating 
the dc-conductance we statistically average over the scattering 
potential assuming that the potential has a vanishing statistical 
mean and that the spectrum of fluctuations is given by the spectrum 
of equilibrium fluctuations. Taking into account interactions in 
the manner outlined above results in  an attenuation of the 
amplitude of AB-oscillations of the statistically averaged 
conductance.
 
In the next section we derive the scattering matrix 
for the MZI. Inelastic scattering from internal potential 
fluctuations is taken into account. In Section \ref{ac} we 
determine the internal potential distribution of the ring 
and then go on to calculate the admittance matrix. We will 
show that by taking into account screening effects a current 
conserving theory for the system consisting of the ring and 
the two gates can be formulated. In the following 
(Section \ref{dcg}) we calculate the dc-conductance and 
investigate the influence of equilibrium fluctuations on dc-transport.

\section{The Mach-Zehnder Interferometer}\label{model}

 We consider an MZI with a single transport channel.  
 An electron arriving at one of the two intersections 
 coming from a reservoir can enter either of the two 
 arms of the ring, but can not be reflected back to 
 a reservoir. An electron coming to the intersection 
 from the ring will enter one of the reservoirs. The 
 amplitudes for going straight through the intersection 
 and for being deflected to the adjacent lead in the 
 forward direction are $t=\sqrt{T}$ and $r=i\sqrt{R}$ 
 respectively, where $T+R=1$. Transmission through the intersections is 
taken to be independent of energy. In the remainder,  
we assume symmetric intersections, that is $R=T=1/2$ in 
(Eq.~\ref{sinel}). Due to the potential 
fluctuations in the arms of the ring a carrier can gain 
or loose 
energy. This process is described by a scattering matrix 
$S_{G}(E',E)$ for each arm which depends on both the 
energy $E$ of the incoming and the energy $E'$ of the 
exiting carrier. The scattering matrix $S_{G}(E',E)$ 
thus connects current amplitudes at a junction of 
the ring incident on the branch 
to the amplitudes of current at the other junction leaving the branch. 
We have a matrix  $S_{G}(E',E)$ for the upper arm ($S_{A}(E',E))$ 
and one for the lower arm ($S_{B}(E',E)$). As a consequence of the 
inelastic transitions in the arms of the 
ring the full scattering matrix  $S_{\alpha\beta}(E',E)$
describing transmission through the entire interferometer from contact 
$\beta$ to contact $\alpha$ is also a function of two energy
arguments. 
This scattering matrix can be found by combining the scattering 
matrices for the two arms with the amplitudes for going through 
the intersections following a specified path. Due to the geometry 
of the system we have the symmetries $S_{13,+B}(E',E)= S_{24,+B}(E',E)$ 
and  $S_{32,+B}(E',E)= S_{41,+B}(E',E)$. In addition the scattering 
matrix elements calculated for the system in  a magnetic field 
$\vec{B}$ are related to the matrix elements found at an inverted 
field $-\vec{B}$ by  $S_{\alpha\beta,+B}(E',E)= S_{\beta\alpha,-B}(E',E)$. 
All elements of the scattering matrix can then be found from the 
three elements given below:  
\begin{eqnarray}\label{sinel}
S_{13}(E',E)&=&S_{24}(E',E)=i\sqrt{TR}\\
&\times &\left(S_A(
E',E){\rm e}^{-i\Phi_A}+S_B(E',E){\rm e}^{+i\Phi_B}\right),\nonumber\\ 
S_{14}(E',E)&=&-R\, S_A(E',E){\rm e}^{-i\Phi_A}+T S_B(E',E) {\rm e}^{+i\Phi_B},\nonumber\\
S_{23}(E',E)&=&T S_A(E',E){\rm e}^{-i\Phi_A}-R\, S_B(E',E) {\rm e}^{+i\Phi_B}.\nonumber
\end{eqnarray} 
Here $\Phi_G$ is the magnetic phase picked up by a particle going 
through arm $G$  
clockwise. Then $\Phi_A+\Phi_B=2\pi\Phi /\Phi_0$, where $\Phi$ 
the flux through the ring and $\Phi_0$ is the flux quantum.
All scattering matrix elements $S_{\alpha\beta}(E',E)$  
with $|\alpha -\beta|\leq 1$ are zero since transport 
through the junctions takes place only in forward direction. 
The on-shell (one-energy) scattering matrix elements for the (free) one-channel 
interferometer in the absence of gates which we 
denote by $S^{(0)}_{\alpha\beta}(E)$ are found by 
replacing $S_G(E',E)$ in Eq.~(\ref{sinel}) by 
$S_G(E',E) = \delta (E' - E) S^{(0)}_G(E)$ with 
$S^{(0)}_G(E) = \exp(ik_EL_G)$. Our first task is now to determine the scattering matrices 
$S_{G}(E',E)$ for the arms of the ring.

\section{S-matrix for a time-dependent potential}\label{HJA}

In this section  we calculate the scattering matrix for the interacting 
ring system. We first solve the Schr\"odinger equation for a single 
branch of the interferometer using a WKB approach. The amplitude for a 
transition from energy $E$ to energy $E+\hw$ of a particle passing through 
this arm and the corresponding scattering matrix element are determined. 
Subsequently the scattering matrices for the two arms are included into a 
scattering matrix for the full interferometer. A  WKB approach similar 
to ours has been used 
previously to discuss photon-assisted transport in a quantum 
point contact \cite{hekking} or to the investigation of traversal 
times for tunneling \cite{landauer}. The influence of a time-dependent 
bosonic environment on transport through a QPC was addressed in 
Refs.~\onlinecite{nazarov} and \onlinecite{schoen} also 
applying a WKB ansatz.

The gate situated opposite to arm $G$ ($G=A,B$) is assumed to be extended over the whole length of this arm. Fluctuations of the charge in the gate capacitively couple to the charge in the neighboring arm of the MZI and influence electron transport through this arm. This interaction effect is taken care of by introducing a time-dependent potential $V_G(x,t)$ into the Hamiltonian 
\begin{equation}\label{hamiltonian}
H = -\frac{\hbar^2}{2m^\ast}\frac{\partial^2}{\partial x^2}+E_G+V_G(x,t),
\end{equation}
for  arm $G$. Here $E_G$ is the sub-band energy due to the lateral confining potential of the arm and $m^\ast$ is the effective mass of the electron.  We make the assumption that the fluctuating potential factorizes in a space- and a time-dependent part, writing $V_G(x,t)=h_G(x)\, e\, U_G(t)$. 
For the ballistic structure considered here the internal potential 
is a slowly varying function of $x$. For practical calculations 
we will however often employ a rectangular potential barrier, ($h_G(x)=const$ if $0\leq x\leq L_G$, where $L_G$ is the length of arm $G$). 
Using  a space-independent internal potential is a valid 
approximation \cite{yaroslav2} 
at least in the low frequency limit $\omega\tau_G \lesssim 1$ where a passing electron sees a constant or slowly changing barrier. 
We have here introduced the traversal time $\tau_G=L_G/v_{G,F}$ where $v_{G,F}$ is the Fermi velocity in arm $G$. We will show  in Section \ref{ac} how the potentials
\begin{equation}\label{fourier}
U_G(t)=\int\,\frac{d\omega}{2\pi}u_G(\omega){\rm e}^{-i\omega t}.
\end{equation}
and their spectra can be determined in a self-consistent way. To solve the Schr\"odinger equation with the Hamiltonian Eq.~(\ref{hamiltonian}) we make the ansatz
\begin{equation}\label{WKBwf}
 \Psi^G_{E}(x,t)={\rm e}^{-iEt/\hbar+i k_{G,E} x+ir_G(x,t)/\hbar},
\end{equation}
 where $k_{G,E}=\sqrt{ 2 m^\ast \left( E-E_G\right) }/\hbar $
 and $r_G(x,t)$ is the action due to inelastic scattering. 
 We will omit the index $G$ of the wave vector (or of the velocity $v_{E,G}=\hbar k_{G,E}/m^\ast$) from now on and only write it when the distinction between wave vectors in different arms is important. 
 It is assumed that transmission is perfect: the potential fluctuations
 cause only forward scattering. 
 This  is justified if all energies relating 
 to the fluctuating potential are much smaller 
 than the Fermi energy $E_F$, in particular  $\hw\ll E_F$.     

In determining $r_G(x,t)$ we will take into account corrections up to the second order in the potential. We write  $r_G(x,t)=r_{G,1}(x,t)+r_{G,2}(x,t)$, where
\begin{equation}\label{r1}
 r_{G,1}(x,t)=\int \,\frac{d\omega}{2\pi} {\rm e}^{-i\omega t}r_{G,1}(x,\omega)
\end{equation}
  is linear in the perturbing potential and 
\begin{equation}\label{r2}
r_{G,2}(x,t)=\int \,\frac{d\omega_1}{2\pi}\int \,\frac{d\omega_2}{2\pi}{\rm e}^{-i\left(\omega_1+\omega_2\right)t}r_{G,2}(x,\omega_1,\omega_2)
\end{equation}
 is a second order correction. The  linear term was calculated 
 in Ref.~\onlinecite{landauer} for a general form of $h_G(x)$. The corresponding expression for the term 
 quadratic in the potential is readily found but is quite cumbersome. 
 We here give $r_{G,1}(x,\omega)$ and $r_{G,2}(x,\omega_1,\omega_2)$ 
 for the case were $h_G(x)$ is a rectangular barrier of length $L_G$:
\begin{eqnarray}\label{r1r2}
r_{G,1}(x,\omega)&=&i\frac{u_G(\omega)}{\omega}\left( {\rm e}^{ix\omega/v_{F}}-1\right),\\
r_{G,2}(x,\omega_1,\omega_2)&=&-\frac{x}{2m^\ast v^3_{F}}\, u_G(\omega_1)u_G(\omega_2){\rm e}^{ix\left(\omega_1+\omega_2\right)/v_{F}}.\nonumber 
\end{eqnarray}  
 Here $r_{G,1}$ gives the contribution to the action due to 
 absorption or emission of single modulation quantum $\hbar \omega$, while 
 $r_{G,2}$ corresponds to the absorption and emission of two modulation quanta 
 $\hbar\omega_1$ and $\hbar\omega_2$. 
 We now proceed to the formulation of the scattering problem in terms of a scattering matrix with elements of the form $S_G(E',E)$ describing transitions between states at different energies. The amplitude $t_G(E',E)$ for a transition from a state with energy $E$ to a state with energy $E'$ of an electron is found from the boundary condition at  $x=L_G$, $\Psi^G_E(L_G,t)=\chi^{G}_E(L_G,t)$.For the matching we expand the WKB wavefunction  (see Eq.~(\ref{WKBwf})) in $x=L_G$ to the second order in the perturbing potential
\begin{eqnarray}\label{expandWKBwf}
 \Psi^G_{E}(x,t)&=&{\rm e}^{-iEt/\hbar+i k_{G,E} x}\bigg[ 1+\frac{i}{\hbar}r_{G,1}(L_G,t) \\
 &+&\frac{i}{\hbar}r_{G,2}(L_G,t)-\frac{1}{2\hbar^2}r^2_{G,1}(L_G,t)\bigg].\nonumber
\end{eqnarray}
Furthermore the wavefunction at the right-hand side of the barrier
(outside the fluctuating potential region) is  
\begin{equation}\label{righthandwf}
 \chi^{G}_E(x,t)=\int \frac{dE'}{2\pi\hbar}t_G(E',E){\rm e}^{ik_{E'}x-iE't/\hbar}.
\end{equation}
 In principle, 
also the derivatives of the wavefunctions should be matched. 
Here we describe transmission through the fluctuating potential 
region as reflectionless which is accurate up to corrections 
of the order of $\hw/E_F$. To determine the transmitted wave with the 
same accuracy it is sufficient to match amplitudes only. 
The transmission amplitude is found by Fourier transforming the WKB wavefunction Eq.~(\ref{expandWKBwf}) and comparing with Eq.~(\ref{righthandwf}). The transmission amplitude can  be expressed in terms of the phase $r_G(L_G,t)$. To second order in the potential we have 
\begin{eqnarray}\label{trans}
t_G(E',E)&=&{\rm e}^{i(k_{E}-k_{E'})L_G}\Bigg[2\pi\hbar \delta(\varepsilon)+\frac{i}{\hbar}r_{G,1}(L_G,\varepsilon/\hbar)\nonumber \\
&+&\int \frac{d\omega}{2\pi}\left[\frac{i}{\hbar}r_{G,2}\left( L_G,\omega,\varepsilon/\hbar-\omega\right)\right.\\ 
&-&\left. \frac{1}{2\hbar^2}r_{G,1}(L_G,\omega)r_{G,1}\left(L_G,\varepsilon/\hbar-\omega\right)\right]\Bigg],\nonumber
\end{eqnarray}
where $\varepsilon =E'-E$. The scattering matrix connecting  incoming wave 
amplitudes (at $x=0$) to outgoing wave amplitudes (at $x=L_G$) is related
 to the transmission amplitude $t_G(E',E)$ through 
\begin{equation}\label{sarm}
S_G(E',E)={\rm e}^{ik_{E'}L_G} t_G(E',E).
\end{equation}
While the transmission amplitude $t_G(E',E)$ was determined through the continuity of the wavefunction in the point $x=L_G$  ($t_G(E',E)$ thus connects amplitudes at the same point), the scattering matrix $S_G(E',E)$ connects amplitudes in $x=0$ to amplitudes in $x=L_G$. This difference in the definitions of the two quantities  leads to the phase factor $\exp (ik_{E'}L_G)$ in Eq.~(\ref{sarm}). The scattering matrix as it is derived here a priori relates wavefunction amplitudes and not current amplitudes. To be consistent with the usual definition  of the scattering matrix as a relation between current amplitudes,  $S_G(E',E)$ should be multiplied by $\sqrt{v_{E'}/v_{E}}$. This factor, however, is of the order $\hw/E_F$ and can thus be neglected. The scattering matrices found to describe a single arm  can now be integrated into the full scattering matrix for the MZI (see Eq.~(\ref{sinel})).
\begin{figure}
\centerline{
\epsfxsize8.5cm
\epsffile{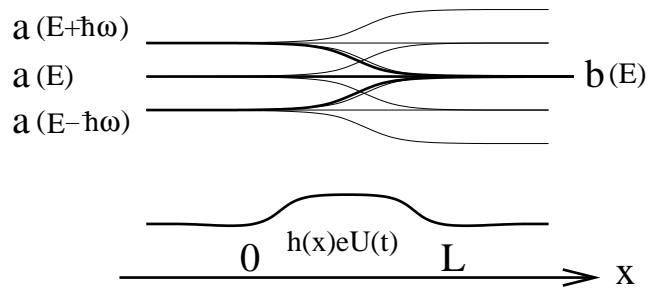}}
\vskip 0.3cm
\caption{\label{potentialfig}
Scattering states $\Psi^G_{E}$ and  $\Psi^G_{E\pm\hw}$ (see Eq.~(\ref{WKBwf})) with amplitudes $a(E)$ and $a(E\pm \hw)$ respectively due to electrons incident from the left at energies $E$ and $E\pm\hw$ are indicated in the figure. For didactic purposes the special case of a harmonically oscillating barrier $U(t)\propto \cos(\omega t)$ is considered and only first order side-bands are drawn. In the rest of the paper we discuss the case of a randomly oscillating barrier and include second order corrections. The scattering state which may be described by a simple (outgoing) plane wave at energy $E$ with amplitude $b(E)$ to the right of the barrier is emphasized in the figure. The amplitudes $a(E)$ and $b(E)$ are related through the scattering matrix via $b(E)=\sum_{\sigma =0,\pm}S(E,E+\sigma \hw)a(E+\sigma \hw)$.} 
\end{figure}

In the discussion presented here the transmission of the carrier 
through the fluctuating potential region is described as a unitary scattering 
process. The ``final'' scattering channels are always open.  
We emphasize that up to now we have investigated a perfectly 
coherent process. Decoherence in our model will be introduced through the statistical averaging (cf. Sec.~\ref{dcg}). 
Our next task is to find the statistical properties of the potential 
fluctuations. These fluctuations can be found from the 
dynamic conductance matrix via the fluctuation-dissipation theorem.

\section{Potential fluctuations}\label{ac}

In this section we proceed to the calculation of the admittance matrix
$G_{lk}(\omega)=dI_{l,\omega}/dV_{k,\omega}$ for the joint system of 
interferometer and gates.  We concentrate on the limit $\hw\ll kT \ll E_F$. 
The dynamic conductance matrix $G_{lk}(\omega)$  is a 
$6\times 6$-matrix ($l,k=1,2,3,4,A,B$), $I_{l,\omega}$ and $V_{k,\omega}$  
denoting respectively the current measured at and the 
voltage applied to one of the four contacts of the ring 
or to one of the two gates. We use the following convention 
for the indices: Lower case Roman indices can take the 
values $ 1,2,3,4,A,B$, the Greek indices $\alpha ,\beta$ 
take the values 1,2,3,4 while upper case Roman 
indices $G,H$ are $A,B$. We will first calculate 
the matrix elements $G_{GG}(\omega)$ from which, 
via the fluctuation dissipation theorem we can 
derive the spectra of potential fluctuations in 
the two arms. These will later be needed in the 
discussion of the decoherence of AB-oscillations. 
The remaining elements of the conductance matrix 
and the resulting  total conductance matrix are 
given in Appendix \ref{admittancematrix}. The 
elements of the conductance matrix  obey the sum-rules 
$\sum_l G_{lk}(\omega)=0$ and $\sum_k G_{lk}(\omega)=0$ 
reflecting gauge invariance and current conservation 
respectively. A problem closely related to the one 
addressed here is concerned with the calculation of 
ac-transport properties of a ballistic wire attached 
to reservoirs and capacitively coupled to a 
gate \cite{yaroslav2}. Contrary to the classical 
calculations done for a wire in Ref.~\onlinecite{yaroslav2} 
the ac-scattering  approach allows us to take 
into account the quantum nature of the system 
investigated here as manifested in the AB-oscillations.

The ac-properties and the potential distribution which are of 
interest here depend not only on the mesoscopic conductor
but also on the properties of the external circuit. 
Here we consider the case where all external current loops
exhibit zero impedance. This requires some explanation since 
especially voltages at the gates are typically controlled
with the help of an external impedance. 
However, what counts in our problem is the 
range of frequencies up to the traversal time, 
whereas the external impedance might be very large only
in a very narrow frequency range around $\omega = 0$. 
Thus we are justified to consider in the following 
a zero-impedance external circuit. 

In order to obtain the conductance matrix from an ac scattering 
approach we need  the effective internal potential $eU_G(t)$ 
in arm $G$. The internal potential $eU_G(t)$  is related to the 
total charge $Q_G(t)$ in the same region through 
$Q_G(t)=C_G(U_G(t)-V_G(t))$ where $V_G(t)$ is the 
voltage applied to gate $G$ and $C_G$ is a geometrical 
capacitance characterizing the strength of the coupling 
between arm and gate. The total charge $Q_G(t)$ consists 
of a contribution due to injection from the contacts 
labeled $Q^e_G(t)$ and a screening part $Q^s_G(t)$, thus $Q_G(t)=Q^e_G(t)-Q^s_G(t)$. 
We will now assume that a voltage $V_{\alpha}(t)$ is 
applied to contact $\alpha$ while  $V_\beta(t)=0$ for $\alpha\neq \beta$. 

First we consider the charge density \cite{ap} injected into 
the arm $G$ due to a modulation of the voltage 
at contact $\alpha$ assuming a fixed internal potential $U_G$. 
The charge distribution in the sample can be expressed through 
the Fermi-field
\begin{equation}
\hat{\Psi}({\bf r},t)=\sum_\alpha\int  \frac{dE}{\sqrt{hv_{\alpha,E}}}\,{\rm e}^{-iEt/\hbar}\psi_\alpha({\bf r};E)\hat{a}_\alpha(E),
\end{equation} 
 which annihilates an electron at point ${\bf r}$ and time $t$. 
 Here $\psi_\alpha({\bf r};E)$ is a 
 scattering state describing carriers with 
 energy $E$ incident from contact $\alpha$. 
 The charge density in the ring at point ${\bf r}$ and time $t$ is $\hat{\rho}({\bf r},t)=e\hat{\Psi}^\dagger({\bf r},t)\hat{\Psi}({\bf r},t)$. Fourier transforming with regard to time and quantum averaging we get $\rho_\alpha ({\bf r},\omega)=\langle\hat{\rho}_\alpha ({\bf r},\omega)\rangle$, where
\begin{eqnarray}\label{density}
\rho_\alpha ({\bf r},\omega)&=&e\sum_{\alpha,\beta}\int\frac{dE}{\sqrt{v_{\alpha ,E}v_{\beta ,E+\hw}}}\\ \nonumber
&\times& \psi^\ast_\alpha({\bf r};E)\psi_\beta({\bf r};E+\hw)\langle\hat{a}^\dagger_\alpha(E)\hat{a}_\beta(E+\hw)\rangle.\nonumber
\end{eqnarray} 
The average charge may be split  into an equilibrium part  $\rho^{(0)}({\bf r},\omega)$ and a contribution due to the external voltage $\delta\rho_\alpha ({\bf r},\omega)$:
\begin{equation}
 \rho ({\bf r},\omega)=\rho^{(0)} ({\bf r},\omega)+\delta\rho_\alpha ({\bf r},\omega).
\end{equation}
 When calculating the quantum average of the charge density operator the effect of the external voltage is taken into account through the modified distribution function for charge carriers coming in from reservoir $\alpha$. The distribution for contact $\alpha$ 
 to linear order in the applied voltage is \cite{ap3} 
\begin{eqnarray}\label{distribution}
\langle\hat{a}^\dagger_\alpha(E)\hat{a}_\alpha(E+\hw)\rangle\\ 
= \delta(\hw)f_\alpha(E)&+&\frac{e}{h}V_{\alpha,\omega}F(E,\omega),\nonumber 
\end{eqnarray}
where $V_{\alpha,\omega}$ is the Fourier component to frequency $\omega$ of the voltage $V_\alpha(t)$  and $F(E,\omega)$ is defined through
\begin{equation}
F(E,\omega)=\frac{f_\alpha(E)-f_\alpha(E+\hw)}{\hw}.
\end{equation}
 Carriers in the other reservoirs are Fermi distributed ($\langle\hat{a}^\dagger_\alpha(E)\hat{a}_\beta(E+\hw)\rangle=\delta(\hw)\delta_{\alpha\beta}f_\alpha(E)$ for $\alpha\neq\mu$ or $\beta\neq\mu$). The scattering states $\psi_\alpha({\bf r};E)$ in the arms of the interferometer for a constant internal potential are of the form $\psi_\alpha({\bf r};E)=A_\alpha\chi({\bf r_\perp})\exp(ik_Ex+i\Phi_G(x))$, where $A_\alpha=i\sqrt{R}$ or $A_\alpha=\sqrt{T}$ depending on the arm and the injecting contact (cf. Eq.~(\ref{sinel})). As in most of the paper we will in the following use $R=T=1/2$. Furthermore  $\Phi_G(x)$ is the magnetic phase acquired going through arm $G$ to point $x$ and  $\chi({\bf r_\perp})$ is the transverse part of the wavefunction. The simple form of the scattering states in the arms is a consequence of the absence of backscattering in the intersections. The injected charge  $\delta\rho_\alpha (x,\omega)$ is the part of the total charge Eq.~(\ref{density}) proportional to the non-equilibrium contribution to the distribution function  Eq.~(\ref{distribution}). Substituting the expressions for the scattering states  into  Eq.~(\ref{density}), using Eq.~(\ref{distribution}) and integrating over ${\bf r_\perp}$ we find 
\begin{eqnarray}\label{density2}
\delta\rho_\alpha (x,\omega)&=&\frac{e^2}{2}\int\frac{dE}{\sqrt{v_{\alpha,E}v_{\alpha,E+\hw}}}\\
&\times& {\rm e}^{i\omega x/v_E}V_{\alpha,\omega}F(E,\omega),\nonumber
\end{eqnarray}
where we have used $|A_\alpha|^2=1/2$.
To find the total charge  $Q^e_{G,\alpha}(\omega)$ injected into  arm $G$ of the MZI we integrate over the length of the arm $ Q^e_{G,\alpha}(\omega)=\int^{L_G}_0 dx \delta\rho_\alpha (x,\omega)$. Performing the integration we get
\begin{equation}
Q^e_{G,\alpha}(\omega)=\frac{e^2}{2h}\int dE\, F(E,\omega)\left(\frac{i}{\omega}\right)\left(1-{\rm e}^{i\omega \tau_G}\right)V_{\alpha,\omega}.
\end{equation}
 In the limit $\hw/kT\ll 1$ we have  $\int dE\, F(E,\omega)\approx1$.  We can rewrite the charge as  $Q^e_{G,\alpha}(\omega)=e^2\nu_{G\alpha}(\omega) V_{\alpha,\omega}$ 
where we have introduced the injectivity $\nu_{G\alpha}(\omega)$, defined as
\begin{equation}\label{injectivity}
 \nu_{G\alpha}(\omega)=\frac{1}{2h}\frac{i}{\omega}\left(1-{\rm e}^{i\omega \tau_G}\right).
\end{equation}
 Here $\tau_G=L_G/v_F$ is the traversal time through arm $G$. 
Now if interactions are taken into account, the excess injected charge will induce a shift in the effective internal potential, which in turn gives rise to a screening charge. This screening charge is proportional to the internal potential $eu_G(\omega)$ and to the total charge density available for screening $\nu_G (\omega)$. Thus  $Q^s_G(\omega)=-e^2\nu_G (\omega) u_G(\omega)$, where $\nu_G (\omega)=\sum^4_{\alpha=1} \nu_{G\alpha} (\omega)=4\nu_{G\alpha} (\omega)$. The last equation is a consequence of the symmetry of the MZI. In the zero frequency limit $\nu_G (\omega)$ reduces to $\nu_G(0)=2L_G/(hv_F)$. The  total charge in region $G$ is $Q_G(\omega)=e^2\nu_G(\omega)(V_{\alpha,\omega}-4u_G(\omega))$.

We generalize now to the case were a voltage is applied not only to one of the contacts but also to gate $G$. The gate voltage is labeled $V_{G}(t)$. In this situation the charge in arm $G$ is  $Q_G(\omega)=C_G(u_G(\omega)-V_{G,\omega})$. Combining with our previous result for the charge leads to
\begin{eqnarray}
Q_G(\omega)&=&C_G(u_G(\omega)-V_{G,\omega})\\
&=&e^2\nu_G(\omega)(V_{\alpha,\omega}-4u_G(\omega)).\nonumber
\end{eqnarray}
Solving for the internal potential and invoking the definition of the injectivity  $\nu_G (\omega)$ (see Eq.~\ref{injectivity}) allows us to express the internal potential $eu_G(\omega)$  through the applied voltages:
\begin{eqnarray}\label{potential}
u_{G}(\omega)=\frac{-i\omega C_G V_{\alpha,\omega}+e^2/(2\, h)\left(1-{\rm e}^{i\omega \tau_G}\right)V_{G,\omega}}{-i\omega C_G +(2\,e^2/ h)\left(1-{\rm e}^{i\omega \tau_G}\right)}.
\end{eqnarray}
The current in gate $G$ is given by $I_{G,\omega}=i\omega Q_G(\omega)$, 
where $-Q_G(\omega)$ is the charge accumulated in the gate. 
Since, with the help of Eq.~(\ref{potential}) 
we can express $Q_G(\omega)$ as a function of external voltages only, 
we can calculate the conductance matrix elements 
$G_{GG} (\omega) = dI_{G, \omega}/dV_{G, \omega}$ and $G_{G\alpha}(\omega)=dI_{G, \omega}/dV_{\alpha ,\omega}$. Note that the matrix elements $ G_{AB}(\omega)$ and $ G_{BA}(\omega)$ vanish since the charge in region $G$ is independent of the voltage applied to the gate further away from it.
(This is a consequence of our assumption of forward 
scattering only at the junctions 
and of the absence of capacitive coupling 
between the two arms). 
For later use we here state the result for  $G_{GG}(\omega)$, which is
\begin{equation}\label{g33}
G_{GG}(\omega)=\frac{dI_{G,\omega}}{dV_{G,\omega}}=\frac{-i\omega C_G}{1-2i\omega C_G R_q/\left(1-{\rm e}^{i\omega \tau_G}\right)}.
\end{equation}
In  Eq.~(\ref{g33}) we have introduced the {\it charge-relaxation} 
resistance $R_q=h/(4e^2)$ of the interferometer. 
The charge relaxation resistance \cite{ap1} is a measure of the 
dissipation generated by the relaxation of excess charge on the conductor 
into the reservoirs. For a structure with perfect 
channels connected to a reservoir each reservoir channel connection 
contributes with a conductance $G_q = 2e^{2}/h$: the conductances 
of different channels add in parallel since each channel reservoir connection 
provides an additional path for charge relaxation. 
For example a ballistic wire connected to two reservoirs has a
charge relaxation resistance $R_q = (G_q  + G_q)^{-1} = h/4e^{2}$.  
For the MZI considered here an excess charge in the upper or lower 
branch has the possibility to relax into the four reservoirs of the MZI. 
But at each junction the two connections are only open with probability 
$T$ and $R$ (see Appendix B). 
Thus the two connections act like one perfect channel. 
As a consequence the charge relaxation resistance for our MZI 
is just that of a perfect wire and also given by $R_q=h/(4e^2)$.
\begin{figure}
\centerline{
\epsfxsize8cm
\epsffile{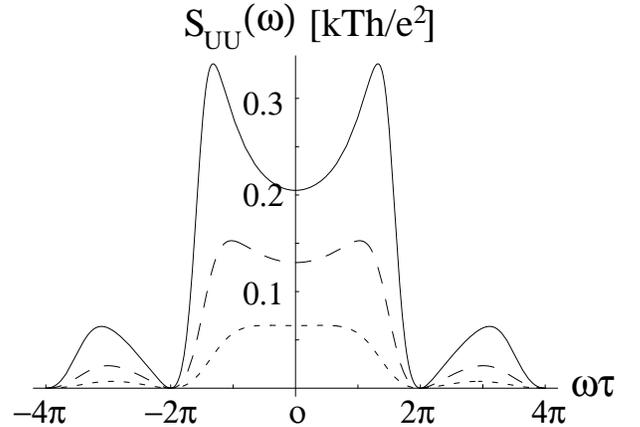}}
\vskip 0.3cm
\caption{\label{3spectra}
 The spectrum Eq.~(\ref{fullspectrum}) is shown as a function 
 of the parameter $\omega\tau$. The full curve corresponds 
 to an  interaction strength $g=0.6$, for the middle (dashed) 
 curve $g=0.7$ and for the lowest (dotted) curve $g=0.8$. } 
\end{figure}
In the low-frequency limit we get from Eq.~(\ref{g33})
\begin{eqnarray}\label{Ggg}
G_{GG}(\omega)&=&- iC_{\mu,G}\omega +R_qC^2_{\mu,G}\omega^2\\
&-&i\frac{1-3g^2_G}{3g_G^2}R^2_qC^3_{\mu,G}\omega^3+\ldots \, .\nonumber
\end{eqnarray}
This is in agreement with the result of Blanter et al.~\cite{yaroslav2} 
for a single wire coupled to a gate. 
We have introduced the dimensionless (Luttinger) 
parameter $g_G$ as a measure of the strength of coupling 
between arm $G$ and gate $G$.  
If arm and gate are decoupled the interaction 
parameter takes the value $g_G=1$ while it 
goes to zero as the strength of coupling is increased.  
The parameter $g_G$ is related to the capacitance $C_G$ and 
to the density of states $D_G=\nu_G(\omega=0)=2L_G/(hv_F)$  
of the wire through \cite{yaroslav2}
\begin{equation}
g^2_G=\frac{1}{1+e^2D_G/C_G}.
\end{equation}
The electrochemical capacitance \cite{ap1} $C_{\mu,G}$ of arm $G$ is 
$C_{\mu,G}^{-1}=C^{-1}_G+(e^2D_G)^{-1}$.

 The remaining elements of the conductance matrix, namely 
those involving the currents in the contacts of the MZI, 
can be derived from an ac-scattering approach. 
These calculations are presented in Appendix \ref{admittancematrix}.
To discuss the influence of potential fluctuations on dc-transport we need  the spectrum  $S_{U_GU_G}(\omega)$ of these fluctuations. Since the spectrum of the current fluctuations $S_{I_GI_G}(\omega)$ in region $G$ is related to the real (dissipative) part of the element $G_{GG}(\omega)$ of the emittance matrix through the fluctuations dissipation theorem $S_{I_GI_G}(\omega)=2kT{\rm Re}\, G_{GG}(\omega)$ (in the limit $\hw\ll kT$), we get $S_{U_GU_G}(\omega)$ from the relation $S_{U_GU_G}(\omega)$=$S_{I_GI_G}(\omega)/(\omega^2C^2_G)$:
\mybeginwide
\begin{equation}\label{fullspectrum}
S_{U_GU_G}(\omega)=kT\frac{h}{e^2}\frac{(1-g_G^2)^2(1-\cos (\omega\tau_G))}{2(1-g^2_G)^2\left(1-\cos (\omega\tau_G)\right)+2g^2_G(1-g^2_G)\omega\tau_G\sin (\omega\tau_G)+g^4_G(\omega\tau_G)^2 }.
\end{equation}
\myendwide
The spectrum Eq.~(\ref{fullspectrum}) is shown in 
Fig.~(\ref{3spectra}) as a function of the dimensionless 
parameter $\omega\tau_G$ for different values of the 
interaction parameter $g_G$. Zeroes of $S_{U_GU_G}(\omega)$ 
occur when $\omega\tau$ is a multiple of $2\pi$. 
This is a consequence of our approximation which 
considers only uniform potential fluctuations. 
\begin{figure}
\centerline{
\epsfxsize8cm
\epsffile{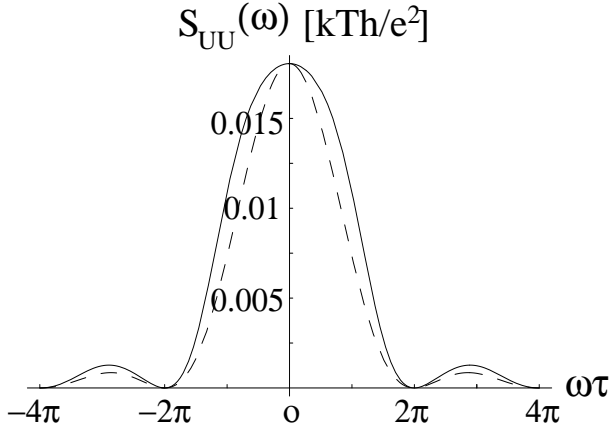}}
\vskip 0.3cm
\caption{\label{trialspectrum}
The exact expression for the spectrum Eq.~(\ref{fullspectrum}) 
(full line) is compared to the expression Eq.~(\ref{spectrum}) 
(dashed line) in the weak coupling limit. 
Here $g=0.9$ is chosen for the interaction parameter.} 
\end{figure}
If the spatial dependence of the potential is taken into  
account, Blanter et al. \cite{yaroslav2} find 
that the traversal time is 
renormalized through the interaction ($\tau_G\rightarrow 
g_G\tau_G$) and consequently the zeroes of  $S_{U_GU_G}(\omega)$ 
are shifted accordingly.
 Instead of the dynamics of single carriers 
it is plasmons which govern the high frequency dynamics. 
This comparison indicates thus the limitation of our approach:
Since we start from a one-particle 
picture our approach is most reliable in the 
case of weak coupling $g_G\rightarrow 1$. In the weak coupling 
limit ($g_G\rightarrow 1$) we expand the spectrum 
Eq.~(\ref{fullspectrum}) to the leading order in  
$C_{\mu,G}/C_G=(1-g^2_G)$ which  leads to 
\begin{equation}\label{spectrum}
S_{U_GU_G}(\omega)= 
2kTR_q\frac{C^2_{\mu,G}}{C^2_G}\frac{\sin (\omega\tau_G/2)^2}{(\omega\tau_G/2)^2}.
\end{equation}
The spectrum vanishes in the non-interacting 
limit $g_G =1$ ($C_{\mu,G} /C_G=0$). In Fig.~\ref{trialspectrum} 
the full spectrum  Eq.~(\ref{fullspectrum}) is compared 
to the approximate form  Eq.~(\ref{spectrum}).
The function ${\sin (\omega\tau_G/2)}/{(\omega\tau_G/2)}$
reflects the ballistic flight of carriers through an interval 
of length $L$.  In the limit of strong coupling the potential noise is 
white and 
\begin{equation}\label{spectrum1}
S_{U_GU_G}(\omega)= 2kT R_q = 2kTh/(4e^2).
\end{equation}
Remarkably in the strong coupling limit 
for the ballistic ring considered here the spectrum 
is universal. The only property of the system which enters is the number 
of leads which permit charge relaxation. 

Having found the fluctuation spectra of the internal potentials we are now in the position to investigate the influence of fluctuations on dc-transport.

\section{Attenuation of the dc-conductance through coupling to a gate}\label{dcg}

We now come to the discussion of the dephasing of the coherent part  
of the dc-conductance in the linear transport regime due to (equilibrium)
 charge fluctuations. In the last section we have shown that 
interactions lead to effective charge and  potential distributions 
in the arms of the ring which in turn give rise to displacement 
currents in the gates and contribute to the ac component of the 
particle currents in the contacts of the ring. The zero frequency 
contribution to the currents in the contacts remained unchanged. 
Here we go one step further and will discuss how  charge fluctuations
 act back on the dc-conductance. In contrast to the last section we
 will thus only discuss the zero frequency component of the conductance.
 Electron-electron interactions affect the coherent part of the 
dc-conductance only. This can be understood from the well 
known result \cite{safi} that interactions do not change the conductance of a one-dimensional 
wire attached to reservoirs. In the interferometer, when interactions 
are considered, AB oscillations of the conductance are suppressed. 
The  dc-conductance matrix for the case without interactions
 is given in appendix \ref{admittancematrix} (see Eq.~(\ref{G0})).
 Throughout this section we choose $\mu_1 =\mu_0+eV$ and
 $\mu_2 =\mu_3=\mu_4=\mu_0$. We will first assume that only arm $A$ 
is coupled to a gate ($C_B\rightarrow \infty$). The generalization
 to the case where both arms are coupled to gates is straight 
forward and will be discussed at the end of this section.  \\
We will from now on treat the potential as a function with certain
 statistical properties. The potential fluctuations are characterized
 by the spectrum $S_{U_AU_A}(\omega)$ which is defined through
\begin{equation}\label{spectrum2}
 2\pi\delta(\omega+\omega ')S_{U_AU_A}(\omega)=\ovl{\langle u_A(\omega)u_A(\omega ')\rangle}.
\end{equation}
The spectrum  was evaluated in Sec.\ref{ac}. In addition the potential has
a vanishing mean value $\ovl{\langle U_A(t) \rangle } =0$. Here  $\ovl{\langle .. \rangle }$ denotes the statistical average and the $u_A(\omega)$ are the Fourier components of $U_A(t)$ (cf. Eq.~(\ref{fourier})). 
The bar is to emphasize the distinction between quantum averages $\langle..\rangle$ and statistical averages. 

The quantity of interest to us is the statistically averaged dc conductance, defined through   
\begin{equation}\label{Gav}
\ovl{\langle G_{\alpha \beta}\rangle}=\lim_{V_\beta\to 0}\ovl{\langle dI_{\alpha}\rangle} /dV_{\beta}.
\end{equation}
A convenient starting point for the calculation of the conductance is the 
following expression for the current in a contact $\alpha$ \cite{mb92},
\begin{eqnarray}\label{current1}
\hat{I}_{\alpha}(t)&=&\frac{e}{h}\int \, dE\, dE'{\rm e}^{i(E-E')t/\hbar}\\
&\times&\left[\hat{a}^{\dagger}_{\alpha}(E)\hat{a}_{\alpha}(E')-\hat{b}^{\dagger}_{\alpha}(E)\hat{b}_{\alpha}(E')\right],\nonumber
\end{eqnarray}
in terms of the operators $\hat{a}^{\dagger}_{\alpha}(E)$ 
$(\hat{a}_{\alpha}(E))$ creating (destroying) an electron in a 
state with energy $E$ entering the system through contact $\alpha$ and
 the operators $\hat{b}^{\dagger}_{\alpha}(E)$ and $\hat{b}_{\alpha}(E)$ 
respectively creating and annihilating an electron outgoing at energy $E$. 
The operators  $\hat{a}_{\alpha}(E)$ and $\hat{b}_{\alpha}(E')$ are related 
through the scattering matrix (see Fig.~(\ref{potentialfig}) and Eq.~(\ref{sinel}))
\begin{equation}\label{smatrix}
\hat{b}_{\alpha}(E)=\sum_{\beta} \int dE'\, S_{\alpha \beta}(E,E') \hat{a}_{\beta}(E').
\end{equation}
 As described in Section \ref{HJA} Eqs.~(\ref{sinel}),(\ref{trans}) and (\ref{sarm}) we determine the scattering matrix elements from the WKB solution of the Schr\"odinger equation for the arm of the ring. 
 Doing this we go to the second order in the perturbing potential. 
This is necessary since due to the assumption of a  vanishing mean 
value of the statistically averaged internal potential there exist 
no first order corrections to the averaged conductance Eq.~(\ref{Gav}). 
Combining Eqs.~(\ref{current1}) and (\ref{smatrix}) with the scattering 
matrix (\ref{sinel}) and statistically averaging leads to the following 
expression for the average conductance: 
\begin{equation}\label{Gdc}
\ovl{\langle G_{\alpha\beta}\rangle} = \frac{e^2}{h}\int\,dE \left(-\frac{\partial f}{\partial E}\right)\ovl{\langle T_{\alpha\beta}(E)\rangle}.
\end{equation}
In above equation we have introduced the statistically averaged 
transmission probability 
\begin{eqnarray}\label{transmission0}
\ovl{\langle T_{\alpha\beta}(E)\rangle} &=& \frac{1}{2}\bigg[1\pm\cos\left(\Theta (E) -2\pi\Phi/\Phi_0\right)\\
&\times& \left(1-\ovl{\langle r^2_{A,1}(L_A,t)\rangle}/2\hbar^2\right) \nonumber\\
&\mp&  \sin\left(\Theta (E) -2\pi\Phi/\Phi_0\right)\ovl{\langle r_{A,2}(L_A,t)\rangle}/\hbar \bigg].\nonumber
\end{eqnarray}
 Here $\Theta =k_{A,E}L_A-k_{B,E}L_B$ is a geometric phase, $\Phi$ is the magnetic flux enclosed by the ring and $\Phi_0$ is the flux quantum.
The upper sign is for the pairs of indices $(\alpha,\beta)=(1,3), (2,4)$, 
the lower sign is for $(\alpha,\beta)=(1,4), (2,3)$. Furthermore, 
$\ovl{\langle T_{\alpha\beta}(E)\rangle}$ and 
$\ovl{\langle T_{\beta\alpha}(E)\rangle}$ are related via the Onsager 
relations.  Expressions for  $r_{A,1}(L_A,t)$ and $r_{A,2}(L_A,t)$ are 
given in Eqs.~(\ref{r1}),~(\ref{r2}) and (\ref{r1r2}). If charge fluctuations
 are not taken into account the transmission probability simply is
 $\ovl{\langle T_{\alpha\beta}(E)\rangle} = 
\left(1\pm\cos\left(\Theta (E) -2\pi\Phi/\Phi_0\right)\right)/2$ 
(compare Eq.~(\ref{G0})). Interactions thus decrease the amplitude
 of the AB-oscillations and lead to an additional out-of-phase contribution.
 Eq.~(\ref{transmission0}) can be rewritten in an approximate 
but more convenient form as 
\begin{eqnarray}\label{transmission1}
\ovl{\langle T_{\alpha\beta}(E)\rangle} &=& \frac{1}{2}\left[1\pm{\rm e}^{-\ovl{\langle r^2_{A,1}(L_A,t)\rangle}/(2\hbar^2)}\right.\\
&\times& \left.\cos\left(\Theta (E) -2\pi\Phi/\Phi_0+\ovl{\langle r_{A,2}(L_A,t)\rangle}/\hbar\right)\right].\nonumber
\end{eqnarray}
Note that also in the presence of interactions current is conserved and 
the system is gauge invariant. This is reflected in the fact that
 $\sum_\alpha \ovl{\langle G_{\alpha\beta}\rangle}=0$ and 
$\sum_\beta \ovl{\langle G_{\alpha\beta}\rangle}=0$ 
(with $\ovl{\langle G_{\alpha\alpha}\rangle}=-e^2/\hbar$ ). 
Eq.~(\ref{transmission1}) has a rather intuitive interpretation 
since 
$\ovl{\langle T_{\beta\alpha}(E)\rangle}\sim 
\ovl{\langle |\Psi^A_E(x,t)+\Psi^B_E(x,t)|^2\rangle}$, 
where $\Psi^A_E(x,t)$ and $\Psi^B_E(x,t)$ 
are the WKB wave-functions for the upper and lower arm respectively 
at energy $E$ (see Eq.~(\ref{WKBwf})).
Note, that Eq.~(\ref{transmission1}) is strictly correct only to second 
order in the fluctuating potential. 

The transmission probability  $\ovl{\langle T_{\alpha\beta}(E)\rangle}$ depends on energy $E$ only through the geometric phase $\Theta =k_{A,E}L_A-k_{B,E}L_B\approx -k_{A,E}\Delta L+\Delta E L_A/(\hbar v_F)$. Here it is assumed that both the difference in length of the two arms $\Delta L=L_A-L_B$ and the difference of the sub-band energies in the two branches of the ring $\Delta E=E_A-E_B$ are small.   The AB oscillations in the conductance are washed out completely by thermal averaging when $\Theta(E+kT)-\Theta(E)\approx 2\pi$. It follows that oscillations may be observed at temperatures $kT\ll E_F(\lambda_F/\Delta L)$. Assuming that $\Delta L$ is of the order of the Fermi-wavelength $\lambda_F$ we conclude that  $kT\ll E_F$ is sufficient  for neglecting the influence of thermal averaging on 
the phase-coherence of the AB-oscillations in the conductance. This rather 
surprising result is a consequence of the absence of 
closed orbits in the interferometer.
It is generally expected that mesoscopic phase coherence 
is destroyed at temperatures of the order of the Thouless energy 
$E_T=E_F(\lambda_F/L)\ll E_F$. If closed orbits were considered 
the transmission probability 
would be a function not only of $k_{E}\Delta L$ but also of $k_E L$. 
However, in our simple model we can write  
$\ovl{\langle G_{\alpha\beta}\rangle}\approx 
(e^2/h)\ovl{\langle T_{\alpha\beta}(E_F)\rangle}$ 
for $kT\ll E_F$ (but still in the limit
$\hw \ll kT$) when assuming that $\Delta L\sim\lambda_F$.   
\begin{figure}
\centerline{
\epsfxsize8.5cm
\epsffile{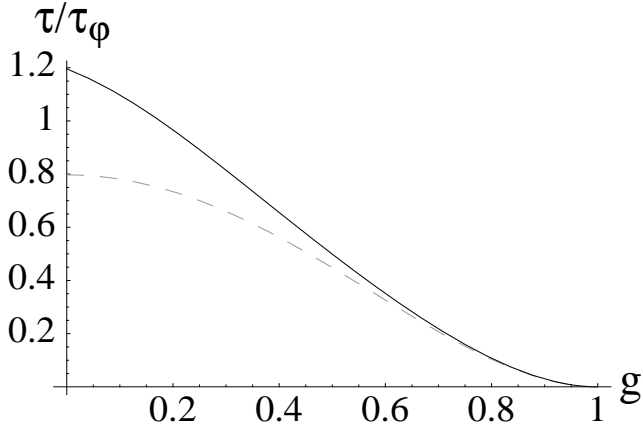}}
\vskip 0.3cm
\caption{\label{detime}
The ratio of traversal time and dephasing time  $\tau/\tau_\phi$ 
versus the coupling parameter $g$. 
The upper (full) line shows the numerically evaluated 
exact result (cf. Eq.~(\ref{dephasing3})), while the 
dashed line is the weak coupling approximation 
Eq.~(\ref{dephasing2}). In the calculations we used 
the parameters of Ref.~[8], 
namely $L_A \sim 1.5 \mu $m, $k_F \sim 1.5\cdot 10^8 {\rm m}^{-1}$.  
Furthermore we assumed $T=1$ K and for the effective mass 
$m^\ast$ of the electron we used the value for GaAs, 
$m^\ast=0.067m_e$.  With these parameters the Fermi 
velocity is $v_F = \hbar k_F/m^\ast \sim 2.6\cdot 10^5 $ m/s. } 
\end{figure}
We can now proceed to further evaluate Eq.~(\ref{transmission1}). Using $r_{A,1}(L_A,t)$ as given in Eq.~(\ref{r1r2}) and invoking the definition of the spectrum  Eq.~(\ref{spectrum2}) we obtain
\begin{equation}\label{dephasing1}
\ovl{\langle r^2_{A,1}(L,t)\rangle}=4e^2\int \frac{d\omega}{2\pi} S_{U_AU_A}(\omega)\frac{\sin^2(\omega\tau /2)}{\omega^2}.
\end{equation}
The fluctuation spectrum $S_{U_AU_A}(\omega)$ has been calculated in Sec.~\ref{ac}. The integral in Eq.~(\ref{dephasing1}) can be done analytically in the two limiting cases of very strong ($g\rightarrow 0$) and weak ($g\rightarrow 1$) coupling between arm and gate. Since the spectrum $S_{U_AU_A}(\omega)$ (Eq.~(\ref{fullspectrum})) is a function of $\omega$ only through $\omega\tau$, it can be seen from  Eq.~(\ref{dephasing1}) that we can then write
\begin{equation}\label{dephasing3}
 \ovl{\langle r^2_{A,1}(L_A,t)\rangle}/(2\hbar^2) =\tau_A/\tau_\phi,
\end{equation}    
where $\tau_A$ is the traversal time and $\tau_\phi$ is a 
function of temperature and the coupling parameter only. 
Eq.~(\ref{dephasing3}) defines the dephasing time $\tau_\phi$. 
It is  expressed through the phase-shift $r_{A,1}(L_A,t)$ acquired by a 
WKB wavefunction in the presence of a time dependent 
potential (relative to the case without potential) and quantifies the strength of the suppression of the AB oscillations in the dc-conductance (see Eq.~(\ref{transmission1})).
In the limit where gate and ring are weakly coupled 
we use the approximate spectrum Eq.~(\ref{fullspectrum}) 
to evaluate Eq.~(\ref{dephasing1}). 
The  dephasing rate $\Gamma_\phi=\tau^{-1}_\phi$ is found to be  
\begin{equation}\label{dephasing2}
\Gamma_\phi= \left(\frac{\pi}{3}\right)\frac{kT}{\hbar}(1-g^2_A)^2 =  \left(\frac{2\,e^2}{3\,\hbar^2}\right)\frac{C^2_{\mu ,A}}{C^2_A}kT R_q.
\end{equation}
The dephasing rate $\Gamma_\phi=\tau^{-1}_\phi$ is linear in temperature. 
Very recently experimental results were reported by Hansen et al. \cite{hansen} on the temperature dependence of decoherence of AB-oscillations in ballistic rings.  In these measurements a dephasing rate linear in temperature was found. The dephasing rate $\Gamma_\phi$  (Eq.~(\ref{dephasing2})) also depends on the coupling parameter $g_A$. Dephasing  goes to zero when gate and ring are completely decoupled ($g_A\rightarrow1$).
 We can similarly determine the dephasing rate in the strong coupling limit. We know that in this case the potential noise is white and the spectrum is given by $S_{U_AU_A}(\omega)=2kTR_q$. The dephasing rate is $\Gamma_\phi=R_q(e^2/\hbar^2)kT$. Due to the more complicated form of the fluctuation spectrum $S_{U_AU_A}(\omega)$ in the intermediate parameter range for $g$ we can't give a simple analytical expression for the dephasing rate. However, the integral in Eq.~(\ref{dephasing1}) can be performed numerically. In Fig.~(\ref{detime})  $\tau/\tau_\phi$ is plotted versus the interaction parameter $g_A$ over the full parameter range.

Returning to the weak coupling limit we further evaluate  $r_{A,2}(L_A,t)$
and find 
\begin{equation}
\delta\Theta=\ovl{\langle r_{A,2}(L_A,t)\rangle}/\hbar=-\frac{\pi}{4}\frac{C^2_{\mu,A}}{C_A^2}\left(\frac{kT}{E_F}\right).
\end{equation}
To be consistent with our previous approximations this term should be neglected since it is of the order $kT/E_F$. Still it is interesting to compare the size of the two corrections due to scattering from the internal potential. We find that $\left(\tau/\tau_\phi\right)/\delta\Theta\sim k_FL\gg 1$ which implies that taking scattering into account to the first order in the potential is a surprisingly good approximation. Combining the information gathered so far allows us to rewrite the transmission matrix elements Eq.~(\ref{transmission1}) in the more convenient form
\begin{equation}\label{transmission2}
\ovl{\langle T_{\alpha\beta}(E)\rangle} = \frac{1}{2}\left(1\pm{\rm e}^{-\tau/\tau_{\phi}}\, \cos \left(\Theta (E)-2\pi\Phi/\Phi_0)\right)\right),
\end{equation}
where the energy dependent part of $\Theta (E)$ is of the order $kT/E_F$ in the limit where the two arms have similar length $\Delta L\sim \lambda_F$. The theory developed so far can be readily adapted to the case where both arms are coupled to gates ($g_G\neq 0, G=A,B$) by making the replacement
\begin{equation}\label{twoarms}
\frac{\tau_A}{\tau_\phi}\rightarrow\left(\frac{\tau_A}{\tau_{\phi ,A}}+\frac{\tau_B}{\tau_{\phi ,B}}\right)
\end{equation}
in Eq.~(\ref{transmission2}).  The simple result  Eq.~(\ref{twoarms}) is an immediate consequence of the fact that potential fluctuations in the two arms are uncorrelated ( $S_{U_AU_B}=S_{U_BU_A}=0$). This can either be seen from the corresponding matrix elements of the admittance matrix (cf. Appendix~\ref{admittancematrix}) or by directly calculating potential correlations, as is done in the low frequency limit in Appendix \ref{chargecorrelations}.

\section{Conclusions}

In this work we have examined dephasing due to electron-electron interactions 
in a simple Mach-Zehnder interferometer (MZI). Without interactions
the MZI exhibits only forward scattering. (However, screening 
will generate displacement currents at all contacts in response 
to a carrier entering the conductor). We have shown how a scattering 
matrix approach can be used to calculate the effect of charge 
fluctuations on the conductance. 
We have first determined 
the internal potentials and their statistical properties in a 
non-perturbative way. 
Subsequently we calculated corrections 
to the dc-conductance up to the second order in the effective 
internal potentials. In the expression for the averaged 
dc-conductance a 
dephasing time $\tau_\phi$ occurs in a 
natural way. It is a measure of the strength 
of the attenuation of the AB-oscillations 
as a function of temperature and coupling 
strength between ring and gates. The dephasing rate 
$\Gamma_\phi=\tau^{-1}_\phi$ is found to be {\it linear} in 
temperature and to depend on the Luttinger coupling parameter  
$g$ through $\Gamma_\phi\propto (1-g^2)^2$, at least 
in the weak coupling limit. Alternatively, it depends 
on the ratio of the electrochemical capacitance $C_{\mu}$ and 
the geometrical gate capacitance $C$ like $(C_{\mu}/C)^{2}$.
In terms of the Coulomb energy $E_{c} = e^{2}/2C$ needed to charge the wire 
and the density of states (inverse level spacing) $D = 2L/hv_F$ this ratio is 
$(1+ 1/(2 D E_c))^{-2}$. 
Such a dependence on $E_c$ cannot be obtained from a Golden rule 
argument in which the coupling between the ring and the gate 
is treated perturbatively. Such a treatment would lead 
to a dephasing rate proportional to $E_{c}^{2}$. 
A dephasing rate proportional to $E_{c}^{2}$ is obtained only in the 
(unrealistic) limit that the level spacing far exceeds the Coulomb energy.
On the other hand we found that the evaluation of the phase accumulated 
during traversal of the conductor is surprisingly well described 
just be the first order perturbation theory in the fluctuating 
potential. 

Recently the temperature dependence of 
AB-oscillations in ballistic rings was investigated 
experimentally by Cass\'e et al. \cite{casse}. 
Since both thermal averaging and dephasing of the 
electronic wavefunctions lead to a decrease in 
the visibility of the AB-oscillations as temperature 
is increased a separation of these different effects 
is of interest. Such an analysis of experimental 
data was carried out by 
Hansen et al. \cite{hansen}. They find that the 
dephasing rate is {\it linear} in temperature 
in agreement with our work.  
The dephasing length 
($l_\phi=v_F\tau$) we have calculated can be of the order 
of the dephasing length observed in this experiment~\cite{hansen} when the 
coupling $g$ is taken to be strong enough. A more detailed comparison 
of our result to the experiment is difficult, since the 
setup of  Ref.~\onlinecite{hansen} is different from MZI 
presented here. In the experimental setup a top-gate is 
used to cover the (two-terminal) AB-ring and no 
side-gates are used. 

We note here only as an aside that a linear temperature dependence 
was also observed in experiments on chaotic cavities \cite{huib}. The theory 
presented here, i.e. the spatially 
uniform fluctuations of the potential in the interior 
of the cavity, will also give rise to a linear in temperature dephasing
rate \cite{anka}. It is also interesting to note 
that our dephasing time shows features similar to the 
inelastic scattering time for a ballistic one-dimensional 
wire. The inelastic scattering time of Ref.~\onlinecite{apel} is inversely 
proportional to temperature and can be written as a simple 
function of the Luttinger liquid parameter measuring the 
strength of electron-electron interactions. Whether the 
inelastic time of Ref.~\onlinecite{apel} is in fact also the dephasing time 
remains to be clarified.

Our discussion has emphasized the close connection 
between the ac-conductance of a mesoscopic sample 
and dephasing. We have given the entire ac-conductance matrix 
for the model system investigated here. A current and charge 
conserving ac-conductance theory requires a self-consistent 
approach to determine the internal potentials and requires 
the evaluation of the displacement currents. 

The displacement currents 
at the gates can in principle be measured. 
Nevertheless the thermal potential fluctuations in the arms of the ring 
do not act as a which path detector \cite{buks}.  
In fact the dephasing rate increases with decreasing capacitance 
and is maximal if $C=0$. In this limit there are no displacement 
currents at the gates. The absence of which path detection 
is reflected in the fact that the charge correlations 
in the two arms of the ring vanish in the equilibrium state of the 
ring. 
  
The work presented here can be extended in many directions. 
Multi-channels systems and systems with backscattering can be considered.
The role of the external circuit can be examined.  
We hope therefore that the work reported here stimulates 
further experimental and theoretical investigations. 

\section*{Acknowledgements}
We thank Yaroslav Blanter for a critical reading of the manuscript.
This work was supported by the Swiss National Science Foundation. 

\appendix

\section{admittance matrix}\label{admittancematrix}

We here give the admittance matrix $G_{lk}(\omega)=dI_{l,\omega}/dV_{k,\omega}$ 
($l,k=1,2,3,4,A,B$) for a symmetric 
($L_A=L_B=L$ and thus  $\tau_A=\tau_B=\tau$) Mach-Zehnder 
interferometer (MZI). 
The results for the asymmetric case are similar but notationally 
more cumbersome. As in the rest of the paper we consider the 
limit $\hw \ll kT$ and $\hw , kT\ll E_F$.
We have shown in Sec.~\ref{ac} how the admittance matrix elements 
that relate the displacement currents in the gates to 
voltages applied at 
a gate or a contact can be calculated. It remains to derive the 
matrix elements that relate currents in the contacts to external 
voltages. To this end we employ the ac-scattering approach 
following Ref.~\onlinecite{ap}. We here give a slightly formalized 
derivation of the results found in Ref.~\onlinecite{ap} and generalize 
the results to accommodate a system like the MZI containing 
several regions described by different internal potentials \cite{curacao} 
(the two arms in the case of the MZI). For recent related work we refer the 
reader to Refs.~\onlinecite{ma,jesus}.  
A  time-domain version of the ac-scattering approach was 
recently introduced in Ref.~\onlinecite{aleiner} to the investigation 
of charge pumping in open quantum dots.

We consider the situation where a time-dependent voltage $V_\mu(t)$ is applied to  contact ($\mu$) of the system. We start from the current operator in the form Eq.~(\ref{current1}). Fourier transforming gives 
\begin{eqnarray}\label{current2}
\hat{I}_{\alpha}(\omega)&=& e\int \, dE\left[\hat{a}^{\dagger}_{\alpha}(E)\hat{a}_{\alpha}(E+\hw)\right.\\
&-&\left.\hat{b}^{\dagger}_{\alpha}(E)\hat{b}_{\alpha}(E+\hw)\right].\nonumber
\end{eqnarray}
To take into account scattering from internal potential fluctuations we use Eq.~(\ref{smatrix}) and write 
\begin{eqnarray}\label{current3}
\hat{I}_{\alpha}(\omega)&=& e\int dE\Bigg[\hat{a}^{\dagger}_{\alpha}(E)\hat{a}_{\alpha}(E+\hw)-\int \frac{dE_1}{2\pi\hbar}\frac{dE_2}{2\pi\hbar}\\
&\times&\sum_{\beta,\gamma}S^\ast_{\alpha \beta}(E,E_1)S_{\alpha \gamma}(E+\hw,E_2)\hat{a}^{\dagger}_{\beta}(E_1)\hat{a}_{\gamma}(E_2)\Bigg].\nonumber
\end{eqnarray}
Our next step is to average this expression quantum mechanically. 
Doing this it has to be taken into account, that the distribution of charge carriers coming in from reservoir $\mu$ is modified due to the time-dependent voltage applied to this contact (see Eq.\ref{distribution}). Since we consider only the linear response we expand the scattering matrix to first order in the internal potentials $u_G(\omega)$. We write
\begin{equation}\label{firstorderS}
S_{\alpha\beta}(E',E)=2\pi\hbar \delta (E-E')S^{(0)}_{\alpha\beta}(E)+S^{(1)}_{\alpha\beta}(E',E),
\end{equation}
where $S^{(0)}_{\alpha\beta}(E)$ is the scattering matrix of the ideal ballistic system (see below Eq.~(\ref{sinel})) and  $S^{(1)}_{\alpha\beta}(E',E)$ is linearly proportional to the perturbing potential. Substituting Eqs.~(\ref{distribution}) and (\ref{firstorderS}) into the current operator  Eq.~(\ref{current3}) and taking the quantum average we see that average current in contact $\alpha$ may be written in the form 
\begin{equation}\label{totalcurrent}
I_{\alpha}(\omega)=\langle\hat{I}_{\alpha}(\omega)\rangle=I^{0}_{\alpha}(\omega)+I^{e}_{\alpha}(\omega)+I^{s}_{\alpha}(\omega).
\end{equation} 
The first term $I^{0}_{\alpha}(\omega)$ in Eq.~(\ref{totalcurrent}) is the dc contribution
\begin{eqnarray}
I^{0}_{\alpha}(\omega)&=&e\delta(\hw)\\
&\times&\sum_\beta\int dE\,f_\beta (E)\left(\delta_{\alpha\beta}-|S^{(0)}_{\alpha\beta}(E)|^2\right).\nonumber
\end{eqnarray}
In the case of interest to us here, $f_\beta (E)=f(E)$ and $I^{0}_{\alpha}(\omega)\equiv 0$ due to the unitarity of the scattering matrix  $S^{(0)}_{\alpha\beta}(E)$. The  current 
\begin{eqnarray}\label{external}
I^{e}_{\alpha}(\omega)&=&\frac{e^2}{h}\int  dE \, \left(\delta_{\alpha\beta} - S^{(0)\ast}_{\alpha\beta}(E)S^{(0)}_{\alpha\beta}(E+\hw)\right)\\
&\times&\frac{f(E)-f(E+\hw)}{\hw}V_{\beta}(\omega),\nonumber
\end{eqnarray}   
 can be understood as the part of the total current directly 
 injected into contact $\alpha$ due to the oscillations of 
 the external potential $V_{\mu}(t)$ (see Ref.~\onlinecite{ap}). 
 The third contribution to the total current $I^{s}_{\alpha\beta}(\omega)$ is the response to the internal potential 
 distribution (compare also  Ref.~\onlinecite{ap})
\begin{eqnarray}\label{internal}
I^{s}_{\alpha}(\omega)&=&-\frac{e^2}{h}\int \, dE \,\left( S^{(0)\ast}_{\alpha\beta}(E)S^{(1)}_{\alpha\beta}(E+\hw,E)\right)\\
&\times&\left( f(E)-f(E+\hw)\right).\nonumber
\end{eqnarray}
We now want to apply the theory developed so far to the calculation of the dynamic conductance of the MZI.  For this example the full scattering matrix is given in Eq.~(\ref{smatrix}). Inelastic transitions are absorbed in the scattering matrices of the arms $S_A(E',E)$ and $S_B(E',E)$. From Eqs.~(\ref{r1r2}), (\ref{trans}) and (\ref{sarm}) we know that to first order in the potential
\begin{eqnarray}\label{sarm2}
S_G(E+\hw,E)&=&2\pi\delta (\omega){\rm e}^{ik_EL_G}\\
&+&{\rm e}^{ik_EL_G}\frac{u_G(\omega)}{\hw}\left(1-{\rm e}^{i\omega \tau_G}\right),\nonumber
\end{eqnarray} 
where we used $k_{E+\hw}\approx k_E+\omega/v$. Expressions for the matrix elements $S^{(1)}(E',E)$ for the interferometer are now easily derived from Eqs.~(\ref{smatrix}) and (\ref{sarm2}). For the calculation of the admittance matrix it is furthermore important to note, that in the limit of interest here ($\hw \ll kT$) Eqs.~(\ref{external}) and (\ref{internal}) can be considerably simplified. The Fermi functions appearing in these equations are expanded to linear order in $\hw /kT$. Since in addition the products of scattering matrix elements contained in Eqs.~(\ref{external}) and (\ref{internal}) do not depend on the energy $E$ but only on the energy difference $E'-E=\hw$ for the scattering matrix used here, the energy integrations can be performed.

 Combining the scattering matrix elements  defined in Eq.~(\ref{sinel}) with the expressions for the currents in the gates $I_G(\omega)=i\omega Q_G(\omega)$ and the currents in the contacts $I_{\alpha}(\omega)=I^{e}_{\alpha}(\omega)+I^{s}_{\alpha}(\omega)$ (see Eqs.~(\ref{external}) and (\ref{internal})) we can now calculate all elements of the conductance matrix. We here consider the case of a perfectly symmetric ($L_A=L_B$) interferometer and give the result to first order in $\omega$. Expanding the admittance matrix in the low frequency limit we can write $G_{lk}(\omega)=G^0_{lk}-i\omega E_{lk}+o(\omega^2)$.  Explicitly, the zeroth order term is
\begin{equation}\label{G0}
{\bf G_0 } = \left( \begin{array}{llllll}

      G_0 & 0 & -G^{+}_\Phi & -G^{-}_\Phi&0&0 \\
     0 &G_0  & -G^{-}_\Phi & -G^{+}_\Phi &0&0 \\
     -G^{+}_\Phi &  -G^{-}_\Phi& G_0 &0 &0&0\\
       -G^{-}_\Phi& -G^{+}_\Phi &0 & G_0&0&0\\
     0&0&0&0&0&0\\
     0&0&0&0&0&0
     \end{array} \right),
 \end{equation}
where we have introduced
\begin{eqnarray}
G_0&=&e^2/h,\\
G^{\pm}_\Phi &=& e^2/(2h)(1\pm\cos(2\pi\Phi/\Phi_0)).
\end{eqnarray}
 Note that in the dc-limit there are no currents in the gates. The first order term $E_{lk}$ is called the emittance matrix. It is of the form 
\begin{equation}
{\bf E } = \left( \begin{array}{llllll}

      -E & -E & E^+_\Phi & E^-_\Phi&-E_{A} & -E_{B} \\
      -E & -E & E^-_\Phi& E^+_\Phi&-E_{A} & -E_{B}  \\
      E^+_\Phi& E^-_\Phi& -E & -E& -E_{A}& -E_{B}  \\
      E^-_\Phi & E^+_\Phi &-E  & -E& -E_{A}& -E_{B} \\
      -E_{A} &-E_{A} &-E_{A} &-E_{A}  & 4E_{A} &0\\ 
      -E_{B} & -E_{B}& -E_{B}& -E_{B} &0 & 4E_{B}
        \end{array} \right).
 \end{equation}
%
%\mybeginwide

%\begin{equation}
%{\bf E } = \left( \begin{array}{llllll}

%      -E & -E & +E_\Phi+\tilde{E } & -E_\Phi+\tilde{E}&-E_{A}/4 & -E_{B}/4 \\
%      -E & -E & -E_\Phi+\tilde{E} & E_\Phi+\tilde{E}&-E_{A}/4 & -E_{B}/4  \\
%      +E_\Phi+\tilde{E}& -E_\Phi+\tilde{E}& -E & -E& -E_{A}/4& -E_{B}/4  \\
%      -E_\Phi+\tilde{E} & +E_\Phi+\tilde{E}&-E  & -E& -E_{A}/4& -E_{B}/4 \\
%      -E_{A}/4 &-E_{A}/4 &-E_{A}/4 &-E_{A}/4  & E_{A} &0\\ 
%      -E_{B}/4 & -E_{B}/4 & -E_{B}/4 & -E_{B}/4 &0 & E_{B}
%        \end{array} \right).
% \end{equation}

%\myendwide
%
The entries of the emittance matrix are defined through
\begin{eqnarray}
E &=&\frac{e^2}{h}\frac{1}{8}\left(C_{\mu ,A}/C_A+C_{\mu ,B}/C_B\right)\tau,\\
E_{G} &=& C_{\mu ,G}/4,\\
E^\pm_\Phi &=&\pm E_\Phi +E+E_{A}/2+E_{B}/2,\\
E_\Phi &=&  \frac{e^2}{2h}\tau\cos (\Phi).
\end{eqnarray}
The electrochemical capacitance $ C^{-1}_{\mu ,G}$ is defined as $ C^{-1}_{\mu ,G}=C^{-1}_G+(e^2\,D_G)^{-1}$, where $D_G=2L_G/(h v_F)$ is the density of states per unit length. The charge relaxation resistance $R_q$ is given  by $R_q=h/(4 e^2)$ and the traversal time is $\tau_G=L_G/v_F$. 
Current conservation implies $\sum^6_{l=1}G_{lk}(\omega)=0$ while as a consequence of gauge invariance $\sum^6_{k=1}G_{lk}(\omega)=0$. Similar sum rules hold for every coefficient in the expansion of $G_{lk}(\omega)$ as a function of $\omega$ (e.g. $\sum^6_{l=1}E_{lk}=0$, $ \sum^6_{k=1}E_{lk}=0$).

\section{Charge-Charge correlations}\label{chargecorrelations}

The spectra of charge fluctuations in the gates or, equivalently, the arms of the interferometer, as well as correlations between the charges in the two gates, can be calculated directly from the knowledge of the scattering matrix $S_{\alpha \beta}$ without first calculating the dynamic conductance as we have done in Sec.~\ref{ac}. This approach which is particularly convenient in the low frequency limit was introduced 
in Ref.~\onlinecite{pedersen} for a mesoscopic structure coupled to a single gate and generalized to the 
case of coupling to more than one gate in Ref.~\onlinecite{am2}. In this section we apply the 
approach of Refs.~\onlinecite{pedersen,am2} to calculate the zero 
frequency limit of the fluctuation spectra of 
the charges in the gates of the MZI. In contrast to the rest 
of this paper we calculate both equilibrium and non-equilibrium 
fluctuations. From Ref.~\onlinecite{am2} it is known that in equilibrium, 
to leading order in frequency, the charge-charge correlations are given by
\begin{equation}\label{SQ}
S^q_{Q_GQ_H}=2kTC_{\mu,G}C_{\mu,H}R^{GH}_q,
\end{equation}  
where $G$ and $H$ specify the gates in the system. Here $C_{\mu,G}^{-1}=C^{-1}_G+(e^2D_G)^{-1}$ is the electrochemical capacitance of gate $G$,  $D_G$ being the density of states. Furthermore $R^{GH}_q$ is the generalized charge relaxation resistance to be introduced below. For $G=H$ Eq.~(\ref{SQ}) gives the spectrum of charge fluctuations in gate $G$ while for $G\neq H$ Eq.~(\ref{SQ}) gives the equilibrium charge correlations between gates $G$ and $H$.  With a small voltage applied to one contact of the system the fluctuation spectra to leading order in the applied 
voltage are  Ref.~\onlinecite{am2}
\begin{equation}\label{SV}
S^V_{Q_GQ_H}=2|eV|C_{\mu,G}C_{\mu,H}R^{GH}_V.
\end{equation}  
The generalized charge relaxation resistance $R^{GH}_q$ and the corresponding quantity in the driven case, the Schottky resistance $R^{GH}_V$ can be 
expressed through the (off-diagonal) elements of a generalized 
Wigner Smith time-delay matrix:
\begin{eqnarray}\label{Res}
R^{GH}_q&=&\frac{h}{2e^2}\frac{\sum_{\alpha \beta} 
D^G_{\alpha \beta}D^{H\ast}_{\alpha \beta}}{D_GD_H},\\
R^{GH}_V&=&\frac{h}{2e^2}\frac{\sum_{\alpha \neq \beta} 
\left(D^G_{\alpha \beta}D^{H\ast}_{\alpha \beta} +
D^G_{\beta \alpha} D^{H\ast}_{\beta \alpha } \right)}{D_GD_H}.
\end{eqnarray}
The density of states $D_G$ in region $G$ is the sum of the 
diagonal matrix elements of the Wigner Smith time-delay matrix, 
$D_G=\sum_\alpha D^G_{\alpha\alpha}$. The elements of the time-delay 
matrix can conveniently be found from the scattering matrix via 
the relation \cite{smith}
\begin{equation}\label{timedelay}
D_{\alpha\gamma}(E)=\frac{1}{2\pi i}
\sum_{\beta}S^\ast_{\beta\alpha}(E)\frac{dS_{\beta\gamma}(E)}{dE}.
\end{equation}
The scattering matrix for the MZI is given in Eq.~(\ref{sinel}). 
We here only need the matrix $S^{(0)}_{\alpha\beta}(E)$ 
which is found from Eq.~(\ref{sinel}) 
by replacing $S_G(E',E)$ in 
Eq.~(\ref{sinel}) by $S_G(E',E) = \delta (E' - E) S^{(0)}_G(E)$ with 
$S^{(0)}_G(E) = \exp(ik_EL_G)$
( here we simply have $\Theta_G=kL_G$). 
To calculate the time-delay matrix in region 
$G$ we replace the energy derivative by a 
derivative with regard to the local potential 
$d/dE \rightarrow -d/dU_G$. Since the scattering 
matrix depends on energy only through the phase 
factors $\Theta_G=kL_G$ it is easy to show \cite{am} that
\begin{equation}\label{phase}
\frac{dS_{\beta\gamma}}{dU_G}= - \pi D_G\frac{dS_{\beta\gamma}}{d{\Theta_G}}.
\end{equation}
We can now use Eqs.~(\ref{Res}) to (\ref{phase}) to calculate 
the generalized charge relaxation and Schottky resistances:
\begin{eqnarray}
R^{AA}_q=R^{BB}_q=\frac{h}{4e^2},\\ 
R^{AB}_q=R^{BA}_q=0,\\ 
R^{AA}_V=R^{BB}_V=\frac{h}{16e^2},\\ 
R^{AB}_q=R^{BA}_q=-\frac{h}{16e^2}.
\end{eqnarray}
Combining these equations with  Eqs.~(\ref{SQ}) and (\ref{SV}) 
we get charge-charge correlations for the equilibrium and 
out-of-equilibrium situations. It is interesting to note 
that in equilibrium the charges in the two gates are {\it uncorrelated}. 
This is a consequence of the absence of closed electronic orbits 
in the ring and the fact that we have 
not introduced a Coulomb coupling between the 
two branches of the ring. For the same reason correlations are independent 
of the magnetic field. This implies, that despite the 
fact that AB-oscillations are observed in the currents measured 
at one of the contacts, the interior of the ring behaves like a 
classical system.  In contrast, the charge fluctuations generated 
by the shot noise are correlated. Like the equilibrium charge fluctuations
they are for the geometry investigated here independent of the AB-flux.

 \end{multicols}
 

\begin{thebibliography}{99}
\bibitem{ap3}
This may be derived from Eq.~(2) in Ref.\onlinecite{ap}.
\end{thebibliography}
\end{document}